\documentclass[epjST]{svjour}
\usepackage{graphics}
\usepackage{amssymb,amsmath,graphicx,cite}
\usepackage{comment}
\newcommand{\be}{\begin{equation}}
\newcommand{\ee}{\end{equation}}
\newcommand{\fn}{\footnote}
\newcommand{\LMS}{\Lambda_{\overline{\rm MS}}}

\newcommand{\lt}{\left}
\newcommand{\rt}{\right}

\newcommand{\non}{\nonumber \\}

\begin{document}
\title{Renormalons in static QCD potential: review and some updates}
\author{Hiromasa Takaura\inst{1}\fnmsep\thanks{\email{htakaura@post.kek.jp}}}
\institute{KEK Theory Center, Tsukuba 305--0801,  Japan}
\abstract{
We give a brief review of the current understanding of renormalons of the static QCD potential
in coordinate and momentum spaces. We also reconsider estimate of 
the normalization constant of the $u=3/2$ renormalon and
propose a new way to improve the estimate.
} 
\maketitle
\section{Introduction}
\label{intro}
The static QCD potential is an essential quantity for understanding the QCD dynamics, 
and at the same time it is suitable to understand renormalon of perturbative QCD.
This is due to the following reasons.
First, it is practically possible to observe renormalon in the perturbative series of the static QCD potential
since it exhibits renormalon divergence at quite early stage, say at NLO.
This is caused by the $u=1/2$ renormalon, which is a very close singularity to the origin of the Borel $u$-plane.
Secondly, the perturbative series is known up to $\mathcal{O}(\alpha_s^4)$ 
\cite{Appelquist:1977tw,Fischler:1977yf,Peter:1996ig,Peter:1997me,Schroder:1998vy,Smirnov:2008pn,Anzai:2009tm,Smirnov:2009fh,Lee:2016cgz}.
This is the highest order that has been reached so far for physical observables. 
The explicit large-order coefficients are helpful to examine
if the perturbative coefficients indeed follow the theoretically expected asymptotic form.
Actually theoretical arguments already revealed detailed asymptotic behaviors of
the perturbative coefficients caused by the renormalon at $u=1/2$ and also 
that at $u=3/2$.

In this paper we first give a review of the current theoretical understanding
of the renormalons in the static QCD potential.
We discuss it both in coordinate space and momentum space,
where totally different features are found.
In particular, we explain a simple formula, presented recently,
to analyze renormalons in momentum space.
Secondly, we move on to discussion on estimation of
normalization constants of renormalons.
Normalization constants are the only parameter which cannot be determined by the
current theoretical argument. One needs to know it 
to subtract renormalons in some methods \cite{Lee:2002sn,Lee:2003hh,Ayala:2019uaw,Takaura:2020byt}.
In this paper, we perform a detailed test on methods to extract normalization constants.
This aims at reconsidering the conclusions in Ref.~\cite{Sumino:2020mxk} and Ref.~\cite{Ayala:2020odx};
Ref.~\cite{Sumino:2020mxk} concluded that the normalization constant of the $u=3/2$ renormalon
cannot be estimated reliably with the NNNLO perturbative series
while Ref.~\cite{Ayala:2020odx} stated that it is possible and estimated the normalization constant from the same series.
Since this difference mainly stems from the difference in analysis method,
we examine validity of different methods.
After this examination, we propose a new way to improve the estimate;
we propose to use the scale consistent with the scaling behavior of asymptotic form of perturbative coefficients,
instead of minimal sensitivity scale. This is a new proposal in this paper.
Finally we give conclusions and supplementary discussion.
In Appendix we summarize the notation used here and 
basic relations to discuss renormalons.

\section{Renormalons in coordinate space}
\label{sec:2}
The first IR renormalon of the static QCD potential is located at $t=1/(2 b_0)$  (or $u_*=1/2$),
which is called the $u=1/2$ renormalon \cite{Aglietti:1995tg}. Here $b_0=(11-2n_f/3)/(4 \pi)$ is the first coefficient of the beta function,
where $n_f$ is the number of quark flavors. See Appendix for our notation, 
where the meaning of parameters $t$ and $u_*$ is explained.
This induces the $\mathcal{O}(r^0)$ renormalon uncertainty to $V_{\rm QCD}(r)$.
The important feature of this renormalon is that it is cancelled in the total energy of the
heavy quark and anti-quark system \cite{Pineda:1998id,Hoang:1998nz,Beneke:1998rk},
\be
E= V_{\rm QCD}(r)+ 2m_{\rm pole} ,
\ee
once the heavy quark  pole mass $m_{\rm pole}$ is expanded perturbatively in terms of a
short distance mass.
Considering analogy to the multipole expansion in classical electrodynamics,
one can understand this cancellation as a consequence of the fact that
the $\mathcal{O}(r^0)$ term couples to the total charge of the system \cite{Sumino:2014qpa, Sumino:2020mxk}.
Since the system is color neutral, there should not be the $\mathcal{O}(r^0)$ term and such an uncertainty.
Once we recognize that the cancellation takes place in the total energy,
we can conclude that the $u=1/2$ renormalon uncertainty of $V_{\rm QCD}(r)$ (that of $m_{\rm pole}$) 
is independent of $r$ ($m_{\rm pole}$). Otherwise, the cancellation does not hold.
Hence the renormalon uncertainty is exactly proportional to the QCD dynamical scale:
\be
{\rm Im} \, V_{\rm QCD}(r)_{\pm}|_{u_*=1/2}=\pm K_{1/2} \Lambda_{\overline{\rm MS}} . \label{onehalf}
\ee
See Appendix for the definition of a renormalon uncertainty.
The constant $K_{1/2}$ is the undetermined parameter in this argument.

The second IR renormalon is considered to be located at $u_*=3/2$
from the study in the large-$\beta_0$ approximation and from the structure of
the multipole expansion in pNRQCD \cite{Aglietti:1995tg,Brambilla:1999xf}. 
The uncertainty is roughly given by $\sim \LMS^3 r^2$.
Recently the detailed structure of the second IR renormalon has been investigated \cite{Sumino:2020mxk,Ayala:2020odx}
within the multipole expansion, which gives the static potential as
\be
V_{\rm QCD}(r)=V_S(r)+\delta E_{\rm US}(r)+\cdots .
\ee
Here $V_S(r)$ is a Wilson coefficient in pNRQCD and identified as 
the perturbative computation of the static potential.
Hence, $V_S(r)$ contains the $u=3/2$ renormalon.
$\delta E_{\rm US}(r)$ is the first non-trivial correction in the $r$ expansion, given by
\be
\delta E_{\rm US}(r)=-i \frac{V_A^2(r;\mu)}{6} 
\int_0^{\infty} dt \, e^{-i \Delta V(r) t}\langle g \vec{r} \cdot \vec{E}^a(t,\vec{0}) \phi_{\rm adj}(t,0)^{ab} g \vec{r} \cdot \vec{E}^b (0,\vec{0})\rangle , \label{deltaEUS}
\ee
whose $r$ dependence is roughly given by $\mathcal{O}(r^2)$.
Here $V_A(r)$ is a Wilson coefficient in pNRQCD and $\Delta V(r) \equiv V_O(r)-V_S(r)$
denotes the difference between the potentials of the octet and singlet states.
Since the $u=3/2$ renormalon uncertainty in $V_S(r)$ is considered to be canceled against 
that of $\delta E_{\rm US}(r)$ \cite{Brambilla:1999xf}, it should have the same $r$-dependence as $\delta E_{\rm US}(r)$.
Hence, we reveal the detailed $r$-dependence of $\delta E_{\rm US}(r)$
to  understand the detailed form of the $u=3/2$ renormalon.  
In eq.~\eqref{deltaEUS}, $r$-dependent quantities are $V_A(r;\mu)$ and $\Delta V(r)$ 
besides the power term, $r^2$.
However, since the IR renormalon in $V_S(r)$ is canceled against 
the UV contribution ($t \sim 0$) of $\delta E_{\rm US}(r)$, we can approximate $e^{-i \Delta V(r) t} \sim 1$
in our present analysis and $\Delta V(r)$ is not relevant here. Therefore
the $u=3/2$ renormalon uncertainty is given by\fn{
If we denote the UV contribution to $\delta E_{\rm US}$
which cancels the IR renormalon by $\delta E_{\rm US}|_{\text{UV contr.}}=V_A^2(r;\mu) r^2 \mathcal{O}(\mu)$,
it should be $\mu$ independent. To obtain eq.~\eqref{threehalf}
we use the fact that
${\rm Im} \, V_S(r)_{\pm}|_{u_*=3/2} \propto V_A^2(r;\mu) r^2 \mathcal{O}(\mu)
=V_A^2(r;\mu_0) r^2 \mathcal{O}(\mu_0)$ and then use eq.~\eqref{RGsolve}.
Note that $\exp[-2 \int_{\alpha_s(\mu_0)}^0 dx \, \gamma(x)/\beta(x)] \mathcal{O}(\mu_0)$
is $\mu_0$ independent.} 
\begin{align}
&{\rm Im} \, V_S(r)_{\pm}|_{u_*=3/2}=\pm K_{3/2} \exp \lt[-2 \int_0^{\alpha_s(r^{-1})} dx \frac{\gamma(x)}{\beta(x)}\rt]V_A^2(r;\mu=r^{-1}) r^2 \LMS^3  \non
&=\pm K_{3/2} [1+\mathcal{O}(\alpha^2_s(r^{-1}))] r^2 \LMS^3  . \label{threehalf}
\end{align}
Here we have solved the RG equation,
\begin{align}
&V_A(r;\mu_0)=\exp \lt[-\int_{\alpha_s(\mu_0)}^{\alpha_s(\mu)} dx \, \frac{\gamma(x)}{\beta(x)} \rt] V_A(r; \mu) \non
&\qquad{} \text{where} \quad{} 
\mu^2 \frac{d V_A(r;\mu)}{d\mu^2}
=\gamma(\alpha_s) V_A(r;\mu)=(\gamma_0 \alpha_s+\gamma_1 \alpha_s^2+\cdots)V_A(r;\mu) \label{RGsolve}
\end{align}
and taken $\mu=r^{-1}$ to show the uncertainty in terms of $\alpha_s(r^{-1})$.
In the last line of eq.~\eqref{threehalf}, we used $\gamma_0=\gamma_1=0$\fn{In Ref.~\cite{Sumino:2020mxk},
we mentioned that $\gamma_1$ is not known, but according to Ref.~\cite{Ayala:2020odx},
it is known to be zero. Here we use it and the correction factor in eq.~\eqref{threehalf} 
now becomes $1+\mathcal{O}(\alpha^2_s)$
although in Ref.~\cite{Sumino:2020mxk} it was $1+\mathcal{O}(\alpha_s)$.}
and $V_A(r)=1+\mathcal{O}(\alpha^2_s(r^{-1}))$. Again $K_{3/2}$ is the undetermined constant.
Although the renormalon uncertainty can be different from $\LMS^3 r^2$,
the correction factor, $1+\mathcal{O}(\alpha_s^2(r^{-1}))$, turns out to be small.

\section{Renormalons in momentum space}
\label{sec:3}
Even though the perturbative series of the static potential in coordinate space 
suffers from seriously divergent behavior, that in momentum space has a good convergence property. 
Recently a simple formula to quantify the renormalon uncertainties of 
the momentum-space potential has been proposed \cite{Sumino:2020mxk}. 
In this formula, one considers  Fourier transform of a coordinate-space renormalon uncertainty.
Since renormalon uncertainties in coordinate space can be revealed systematically 
within the multipole expansion as seen above, it provides us with a clear way to 
study momentum-space renormalon uncertainties.

The momentum-space potential $\alpha_V(q)$ is defined by
\be
- 4 \pi C_F \frac{\alpha_V(q)}{q^2}
=\int d^3 \vec{r}  \, e^{-i \vec{q} \cdot \vec{r}}  V_S(r) . \label{alphaV}
\ee
Let us first consider a renormalon uncertainty of simple form in coordinate space:
\be
{\rm Im} \, v(r)_{\pm}=\pm K_{u_*} (\LMS^2 r^2)^{u_*} ,
\ee
where $v(r):=r V_S (r)$ is the dimensionless potential.
The $u=1/2$ renormalon uncertainty indeed takes this form.
We calculate the corresponding renormalon uncertainty in $\alpha_V(q)$ 
by considering Fourier transform of the above renormalon uncertainty.
In other words, we replace $V_S(r)$ in eq.~\eqref{alphaV} with ${\rm Im} \, [v(r)/r]_{\pm}$
to obtain ${\rm Im} \, \alpha_V(q)_{\pm}$.
We obtain
\be
{\rm Im} \, \alpha_V(q)_{\pm}=\mp \frac{K_{u_*}}{C_F} \lt(\frac{\LMS^2}{q^2} \rt)^{u_*} \Gamma(2u_*+1) \cos(\pi u_*) .
\ee
If $u_*$ is a positive half-integer, this uncertainty completely vanishes
since $\cos(\pi u_*)=0$ and $\Gamma(2 u_*+1)$ is finite.
Hence, we conclude that the $u=1/2$ renormalon is absent in the momentum-space potential.
This is a revisit of the old conclusion obtained in Ref.~\cite{Beneke:1998rk}.
Our argument does not rely on diagramatic analysis.

We can easily extend this argument to study renormalon structure in momentum space
beyond the $u=1/2$ renormalon.
Since a general renormalon uncertainty may include logarithms $\log(\mu^2 r^2)$
when we rewrite $\alpha_s(r^{-1})$ in terms of $\alpha_s(\mu)$ [for instance, see eq.~\eqref{threehalf}],
we assume that a renormalon uncertainty is given by
\be
{\rm Im} \, v(r)_{\pm}=\pm K_{u_*} (\LMS^2 r^2)^{u_*} \sum_{n \geq 0} a_n \frac{\partial^n}{\partial u^n} (\mu^2 r^2)^u \big|_{u \to 0} ,
\ee
where $a_n$ is a function of $\alpha_s(\mu)$. 
(In the case of the $u=1/2$ renormalon studied above, $a_n=0$ for $n \geq 1$, because
its uncertainty is exactly proportional to $\LMS$ and does not have $\mathcal{O}(\alpha_s(r^{-1}))$ correction.)
Repeating a similar calculation, we obtain the renormalon uncertainty in momentum space 
induced by the above coordinate-space renormalon uncertainty as
\be
{\rm Im} \, \alpha_V(q)_{\pm}
=\mp \frac{K_{u_*}}{C_F}  \lt(\frac{\LMS^2}{q^2} \rt)^{u_*}  \sum_n a_n \frac{\partial^n}{\partial u^n} 
\lt[ \lt(\frac{\mu^2}{q^2} \rt)^u \Gamma(2 (u_*+u)+1) \cos(\pi(u_*+u)) \rt] \bigg|_{u \to 0}.
\ee
Using this formula one can generally study the detailed renormalon structure in momentum space.
In the case of the $u=3/2$ renormalon 
the uncertainty is given by
\begin{align}
&{\rm Im} \, v(r)_{\pm}|_{u_*=3/2}
=\pm K_{3/2}(\mu) V_A(r;\mu) r^3 \LMS^3 \non
&=\pm K_{3/2}(\mu)  (\LMS^2 r^2)^{3/2} \{1+2 e_2 \alpha_s^2(\mu)+2 [e_3+ (2 b_0 e_2+\gamma_2) \log(\mu^2 r^2)] \alpha_s^3(\mu) + \cdots \} ,
\end{align}
where we denote the perturbative series of $V_A(r)$ as 
$V_A(r)=1+e_2 \alpha_s^2(\mu)+[e_3+ (2 b_0 e_2+\gamma_2) \log(\mu^2 r^2)] \alpha_s^3(\mu) + \cdots$;
$e_2$ and $e_3$ are log independent constants.
Here we used $e_1=\gamma_0=\gamma_1=0$.
Then we obtain the $u=3/2$ renormalon uncertainty in momentum space as
\be
{\rm Im} \, \alpha_V(q)_{\pm}|_{u_*=3/2}
=\mp \frac{K_{3/2}(\mu)}{C_F} \lt(\frac{\LMS^2}{q^2} \rt)^{3/2} [12 \pi (2 b_0 e_2+\gamma_2) \alpha_s^3(\mu)+\cdots] .
\ee
We can see that the momentum-space renormalon uncertainty is suppressed by $\mathcal{O}(\alpha_s^3)$.

We saw that in momentum space the $u=1/2$ renormalon is absent and the $u=3/2$ renormalon 
is fairly suppressed.
It means that the renormalons in coordinate space are caused by
the $q \sim 0$ region in the Fourier integral 
\be
V_{S}(r)=-4 \pi C_F \int \frac{d^3 q}{(2\pi)^3} e^{i \vec{q} \cdot \vec{r}} \frac{\alpha_V(q)}{q^2} .
\ee
This is exactly the case for the $u=1/2$ renormalon
and this is the case to a large extent also for the $u=3/2$ renormalon.
Hence, if we introduce an IR cutoff to the Fourier integral,
\be
V_{S}(r;\mu_f)=-4 \pi C_F \int_{q^2 < \mu_f^2} \frac{d^3 q}{(2\pi)^3} e^{i \vec{q} \cdot \vec{r}} \frac{\alpha_V(q)}{q^2} , \label{IRcut}
\ee
it is almost free from the IR renormalons at $u=1/2$ and $u=3/2$.
We note that the absence of the $u=1/2$ renormalon in this quantity was revealed in Ref.~\cite{Beneke:1998rk} 
and this gave a motivation to define the potential subtracted (PS) mass.

\section{Estimates of normalization constant}
\label{sec:4}
The normalization constants of renormalons, $K$, cannot be determined by the above theoretical arguments.
However it can be estimated from fixed order perturbative coefficients
since the size of normalization constants is related to the asymptotic form of perturbative coefficients.  
The following two methods have been adopted in the literature to extract the normalization constant of a leading IR renormalon.
Hereafter we study $N$ instead of $K$; see eq.~\eqref{relation} for their relation.
 
\vspace{1mm} 

{\bf Method A} \cite{Lee:1996yk}\\
From eq.~\eqref{BorelAroundSing}, 
one considers the function, 
\be
(1-b_0 t/u_*)^{1+\nu_{u_*}} B_v(t) /(\mu^2 r^2)^{u_*}.
\ee
Expanding this function in $t$ and then substituting $t \to u_*/b_0$,
one obtains the normalization constant $N_{u_*}$.
Note that the convergence radius of the series expansion of the above function 
is $\rho=u_*/b_0$ and on the convergence radius it gives us the correct value.

\vspace{1mm} 

{\bf Method B} \cite{Bali:2013pla,Ayala:2014yxa,Ayala:2020odx} \\
Since the asymptotic behavior of perturbative coefficients can be predicted except for the overall constant
one can determine the normalization constant by
\be
N_{u_*}=\lim_{n \to \infty} \frac{d_n}{d_n^{u_* \text{(asym)}}/N_{u_*}} .
\ee
$d_n^{u_* \text{(asym)}}$ is given by eq.~\eqref{asymptotic}.

\vspace{1mm}
\noindent
Both methods should give an accurate answer if an all-order perturbative series is known.


It is stated in Refs.~\cite{Bali:2013pla,Ayala:2014yxa,Ayala:2020odx}
that Method B practically shows faster convergence than Method A
and gives more stable estimate against scale variation.
Related to this, different conclusions were obtained in Ref.~\cite{Sumino:2020mxk}
and Ref.~\cite{Ayala:2020odx} about the estimate of 
the $u=3/2$ renormalon normalization constant, $N_{3/2}$.
In Ref.~\cite{Sumino:2020mxk}, using Method A we concluded that we cannot 
reasonably estimate the normalization constant because
a large uncertainty remains with the NNNLO perturbative series.
In Ref.~\cite{Ayala:2020odx}, on the other hand, the authors presented an estimation of 
the normalization constant using Method B from the same order perturbative series.



In this section, we reconsider the conclusion obtained in Ref.~\cite{Sumino:2020mxk}
by examining whether Method B actually gives more accurate results or not. 
We perform a validity test by considering a model-like all order perturbative series
where exact normalization constants are known.
In fact, since there are little cases where the normalization constant can be
exactly known, such a test would be useful in explicitly examining efficiency and accuracy of the estimation methods.
The model-like all order series considered here is, however, 
realistic enough and is beyond the large-$\beta_0$ approximation. 
In this test we also propose a way to improve the estimate.
After this test, we estimate $N_{3/2}$ from the NNNLO perturbative series 
using the improved method.

\subsection{Test of estimation methods using model-like series}
In the test, we use the all-order perturbative series constructed as follows \cite{Sumino:2005cq}.
We consider the N$^k$LO perturbative series in momentum space with $\mu=q$:
\be
\alpha_V(q)=\sum_{m=0}^k d_{m}(\mu=q) \alpha_s(q)^{m+1}  .
\ee
By rewriting $\alpha_s(q)$ in terms of $\alpha_s(\mu)$, 
we obtain an all-order perturbative series in momentum space.
Here we consider the $(k+1)$-loop beta function. 
Next we perform Fourier transform and can obtain an all-order
perturbative series in coordinate space.
The perturbative series in coordinate space possesses renormalon uncertainties of $u=1/2$, $3/2 \dots$. 
We call this perturbative series the N$^k$LL model series.
The normalization constants of these renormalon uncertainties were exactly calculated 
in Ref.~\cite{Sumino:2020mxk} (see eqs.~(6.6) and (6.9) therein).
In the following analysis, we use the N$^3$LL model series.
(We regularize the IR divergence in the three-loop coefficients in Scheme A defined in Ref.~\cite{Sumino:2020mxk}.)

We demonstrate how we can obtain model series in the simplest case, i.e. the LL model series.
In momentum space, we have the LL result,
\be
\alpha_V(q)|_{\rm LL}=\alpha_s(q)=\alpha_s(\mu) \sum_{n=0}^{\infty} [ b_0 \alpha_s(\mu) \log(\mu^2/q^2)]^n .
\ee
This is an all-order series in terms of $\alpha_s(\mu)$ in momentum spcae.
The all-order perturbative series in coordinate space can be obtained by
\begin{align}
V_{S}(r)|_{\rm LL}
&=-4 \pi C_F \int \frac{d^3 q}{(2\pi)^3} e^{i \vec{q} \cdot \vec{r}} \frac{\alpha_V(q)|_{\rm LL}}{q^2} \non
&=-4 \pi C_F \sum_{n=0}^{\infty} \alpha_s(\mu) (b_0 \alpha_s(\mu))^n 
\int \frac{d^3 q}{(2\pi)^3} \frac{e^{i \vec{q} \cdot \vec{r}}}{q^2} \lt[\log{\lt(\frac{\mu^2}{q^2} \rt)} \rt]^n .
\end{align}
From the integration of logarithms in a small-$q$ region,
$V_S(r)|_{\rm LL}$ has renormalons at $u=1/2$, $3/2$,....
The resummed result is given by the right-hand side of the first equality. 
(We deform the integration path to avoid the singularity of $\alpha_V(q)=\alpha_s(q)$.)
In this expression, the renormalon uncertainties stem from the simple pole of $\alpha_s(q)$ at $q=\LMS$.
The normalization constants of the renormalons can be calculated by the contour integral surrounding the pole.
This idea can be generalized for the N$^k$LL case \cite{Sumino:2020mxk}.
We note that, by construction, if one considers the N$^k$LL model series,
the perturbative coefficients up to the $\mathcal{O}(\alpha_s^{k+1})$ order are exactly obtained,
while the higher order coefficients are estimated based on the expectation that 
the logarithmic terms in $\alpha_V(q)$ dominantly determine the perturbative coefficients in $V_S(r)$.

Since we are now interested in the $u=3/2$ renormalon, we consider
the QCD force, $f(r)=r^2 d V_S(r)/dr$ to eliminate the $u=1/2$ renormalon.
We denote its perturbative coefficient by $d_n^f$.
Once we obtain the normalization constant of the force $N_{3/2}^f$
we can readily obtain the normalization constant of the potential by the relation $N_{3/2}^f=2 N_{3/2}^v$.
Hereafter $N_{3/2}$ means $N_{3/2}^v$.
We assume the number of flavors to be $n_f=3$ throughout this analysis.

We explain how we estimate a central value and its error from the $n$th order truncated series
using Method A or B.
We adopt a parallel estimation method to Ref.~\cite{Ayala:2020odx}.
The central value at the $n$th order (which is estimated from the  N$^n$LO perturbative series)
is determined at the minimal sensitivity scale $\mu r$ of a normalization constant.\fn{
If the minimal sensitivity scale is not found in the range $1/2 < \mu r <5$, 
we treat $\mu r=1$ as the minimal sensitivity scale.}
To estimate the error we vary $\mu r$ around the minimal sensitivity scale 
by the factor $\sqrt{2}$ or $1/\sqrt{2}$.
In addition we obtain the $(n-1)$th order result at the minimal sensitivity scale of the $n$th order result
and examine the difference.  
The procedure so far is common to Method A and B.
In Method B we also examine the difference caused by including $1/n$ correction [i.e. $k=1$ term in eq.~\eqref{asymptotic}]
in $d_n^{u_* \text{(asym)}}$ or not. 
Finally combining the two (three) errors in Method A (Method B) in quadrature,\fn{
The final error is estimated in this way in Ref.~\cite{Ayala:2020odx} and we follow it.}
we obtain the $n$th order result with the total error.

\begin{figure}
\begin{minipage}{0.5\hsize}
\begin{center}
\includegraphics[width=6.2cm]{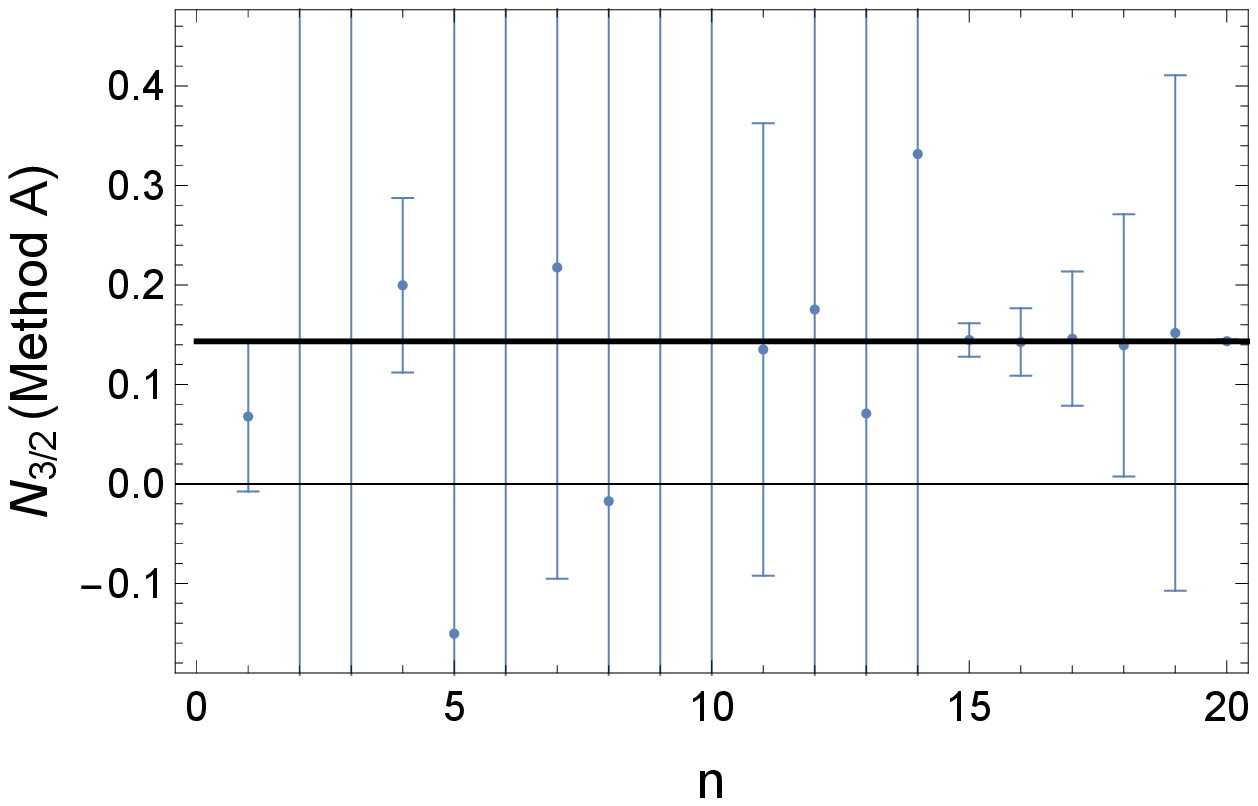}
\end{center}
\end{minipage}
\begin{minipage}{0.5\hsize}
\begin{center}
\includegraphics[width=6.2cm]{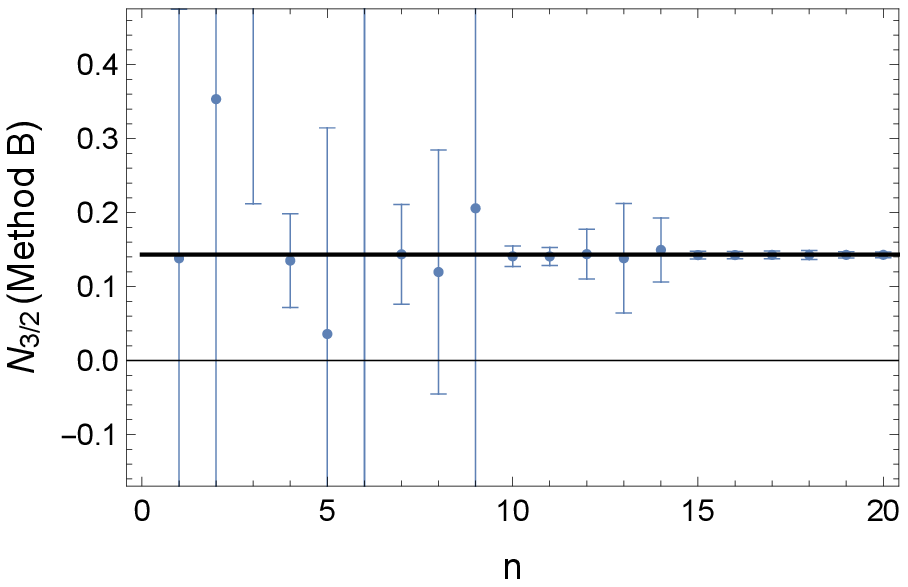}
\end{center}
\end{minipage}
\caption{Estimate of the normalization constant $N_{3/2}$ using Method A (left)
and Method B (right). 
The black line shows the exact answer.}
\label{fig1}
\end{figure}

We show the results in Fig.~\ref{fig1}.
We can see that Method B gives smaller error than Method A
and shows faster convergence.
This indeed agrees with the statement in Refs.~\cite{Bali:2013pla,Ayala:2014yxa,Ayala:2020odx},
and we consider Method B superior.

However, it is worth noting that in both methods 
the error size does not show healthy convergence at small $n$ as seen from Fig.~\ref{fig1};
the estimated error does not always get smaller as $n$ is raised in the region $n \lesssim 10$
in Method B and such a tendency is worse in Method A.

\begin{figure}
\begin{minipage}{0.5\hsize}
\begin{center}
\includegraphics[width=6.2cm]{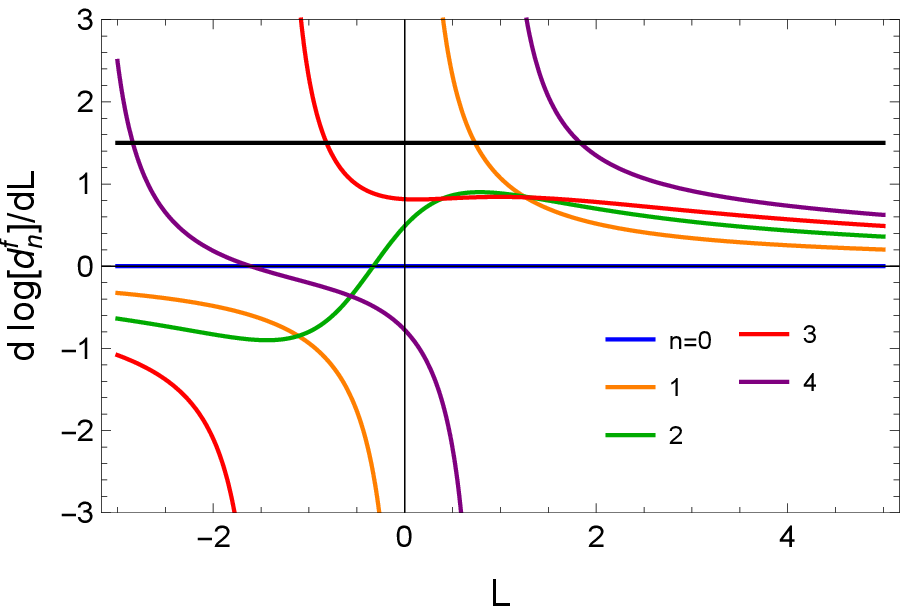}
\end{center}
\end{minipage}
\begin{minipage}{0.5\hsize}
\begin{center}
\includegraphics[width=6.2cm]{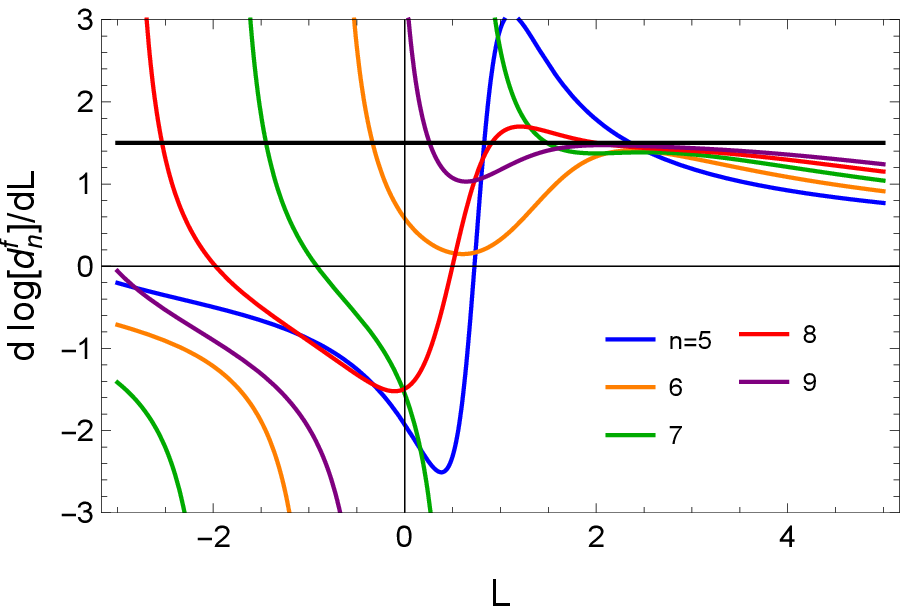}
\end{center}
\end{minipage}
\begin{minipage}{0.5\hsize}
\begin{center}
\includegraphics[width=6.2cm]{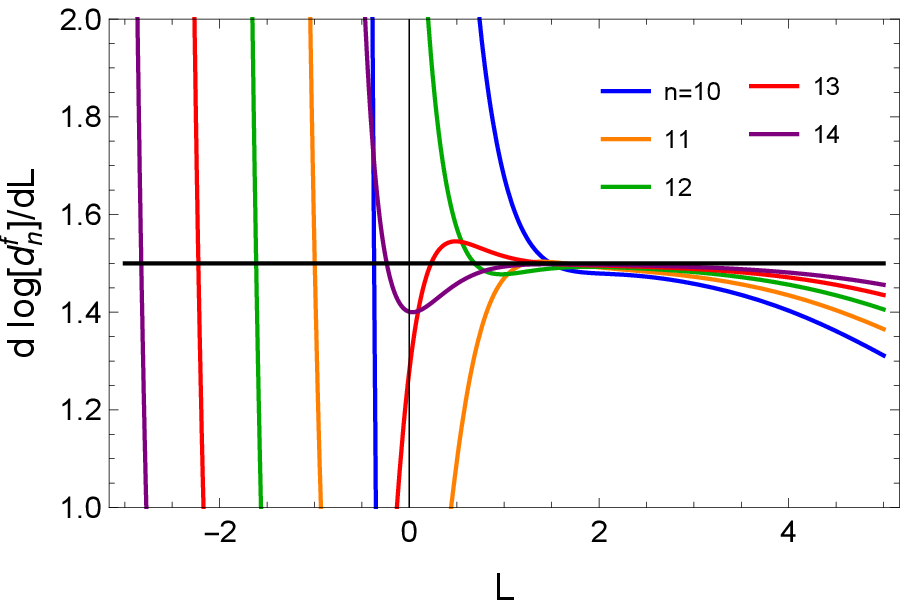}
\end{center}
\end{minipage}
\begin{minipage}{0.5\hsize}
\begin{center}
\includegraphics[width=6.2cm]{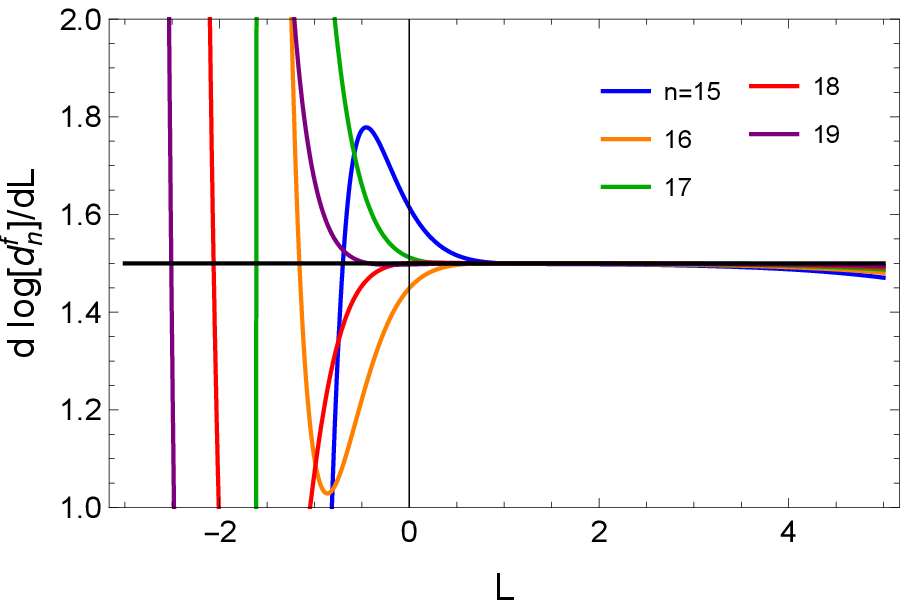}
\end{center}
\end{minipage}
\caption{$d \log(d_n^f)/d L$ for various $n$.
The expected value $3/2$ is shown by the black line in each figure.}
\label{fig2}
\end{figure}

To improve the estimate, we propose
 to use, instead of the minimal sensitivity scale, 
a reasonable scale from the viewpoint of the asymptotic behavior. 
As shown in eq.~\eqref{asymptotic}, $d_n^f$ should behave as
$d_n^f \propto (\mu^2 r^2)^{3/2}$ for scale variation\fn{
Rigorously speaking, ${d_n^f}^{u_* {\rm (asym)}} \propto (\mu^2 r^2)^{u_*}$ does not exactly hold 
in general cases because $c_{k, u_*}$ is a polynomial of $\log(\mu^2 r^2)$.
When a renormalon uncertainty is exactly proportional to $\LMS^{2 u_*}$,
$c_{k, u_*}$ does not have $\log(\mu^2 r^2)$ dependence and $d_n^{u_* {\rm (asym)}} \propto (\mu^2 r^2)^{u_*}$ is exact.
} 
if the $u=3/2$ renormalon dominates the $n$th order perturbative coefficient.
In this case, $d \log{d_n^f}/ d L$ should be (or very close to) $3/2$, where $L=\log(\mu^2 r^2)$.
Showing $d \log{d_n^f}/ d L$ is also useful for checking whether the 
$u=3/2$ renormalon dominates the $n$th perturbative coefficient or not.
We show it in Fig.~\ref{fig2}.
From this figure, we consider that 
the dominance of the $u=3/2$ renormalon sets in around $n \gtrsim 5$.
In the estimate of the normalization constant,
we propose to use the scale where $d \log{d^f_n}/ d L$ is close to $3/2$.
Quantitatively we choose the optimal scale $\mu_0$ such that the integral
\be
\int_{\log( (\mu_0/\sqrt{2})^2 r^2)}^{\log( (\sqrt{2} \mu_0)^2 r^2)} dL \,
\lt|\frac{d \log{d_n^f}}{d L} -\frac{3}{2} \rt|
\ee 
is minimized. Then we determine a central value at $\mu=\mu_0$ using Method B.
We call this estimation method Method B'.
In this method, as seen from Fig.~\ref{fig2}, larger scale $L \sim 2$ is favored,
although in the previous analysis the minimal sensitivity scale appeared $L \sim 0$.
The way to estimate the error is parallel to the previous case;
we examine the difference caused by the scale variation by the factor $\sqrt{2}$ or $1/\sqrt{2}$
and examine the difference of the $(n-1)$th and $n$th order results at $\mu=\mu_0$ .
We also examine the impact of the $1/n$ correction.

\begin{figure}
\begin{center}
\includegraphics[width=8cm]{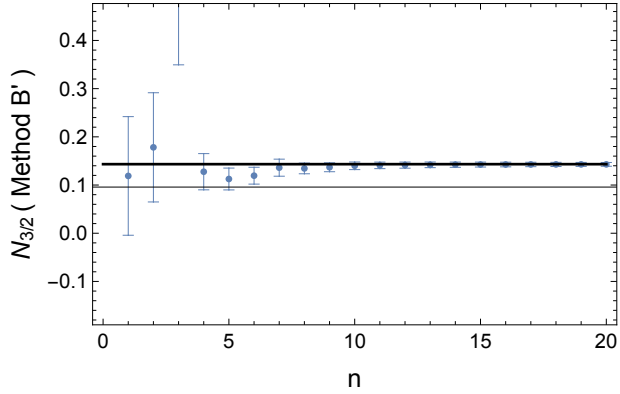}
\end{center}
\caption{Estimate of the normalization constant $N_{3/2}$ using Method B',
where the scale consistent with the $u=3/2$ renormalon is used instead of the minimal sensitivity scale. 
The black line shows the exact answer. The range of the vertical axis is taken the same as Fig.~\ref{fig1}.}
\label{fig3}
\end{figure}

We show the result in Method B' in Fig.~\ref{fig3}.
The convergence is faster than Method B, and remarkably,
the error gets smaller almost monotonically as $n$ is raised, especially at $n \gtrsim 5$.
Hence Method B' is optimal as far as we have tested.
Although the estimate at $n=3$, $N_{3/2}=0.64 \pm 0.29$, deviates from the exact value $0.143$,
this is not surprising because the renormalon would not be relevant enough at this order as suggested from Fig.~\ref{fig2}.

Although we have used the N$^3$LL model series so far, 
we did a parallel analysis using the N$^2$LL model series.
The situation was almost parallel.
We found that (i) Method B shows faster convergence than Method A,
and (ii) Method B' makes the central value and its error converge faster than Method B.

\subsection{NNNLO estimate of $N_{3/2}$}
\begin{figure}
\begin{center}
\includegraphics[width=8cm]{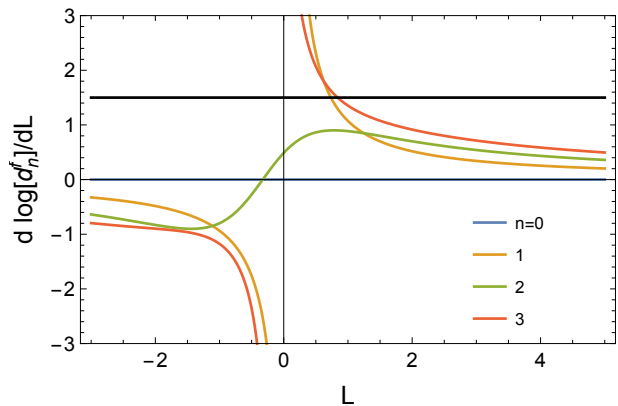}
\end{center}
\caption{$d \log(d_n^f)/d L$ up to NNNLO.
The expected value $3/2$ when the $u=3/2$ renormalon dominates is shown by the black line in each figure.}
\label{fig4}
\end{figure}

We give the NNNLO estimate using Method B'.
So far we have regularized the IR divergence in the three-loop coefficient \cite{Appelquist:1977es,Brambilla:1999qa,Kniehl:1999ud,Brambilla:1999xf}
in Scheme A which is defined in \cite{Sumino:2020mxk}.\fn{
In Scheme A, we assume dimensional regularization in calculating the three-loop coefficient.
Then we drop the divergent term $1/\epsilon$ (associated with the IR divergence) 
and set the renormalization scale to $1/r$.
(Both of the soft and ultra-soft renormalization scales are set to $1/r$.)}
(In this case our estimate at NNNLO reads $N_{3/2}=0.64 \pm 0.29$ as shown above,
although the scaling behavior is far from the expected one from the $u=3/2$ renormalon.)
In this analysis we adopt the same regularization as Ref.~\cite{Ayala:2020odx} to make comparison easy.
We show $d \log d_n^f/d L$ in Fig.~\ref{fig4}.
The difference from the upper left figure in Fig.~\ref{fig2} for $n=3$ comes from 
the difference in regularization of the IR divergence.
In this case, the behavior of $d \log d_3^f/d L$ is closer to $3/2$ than that of Fig.~\ref{fig2}.
We obtain
\be
N_{3/2}=0.17 \pm 0.05_{\text{scale}} \pm 0.02_{\text{NNLO}} \pm 0.02_{1/n} \pm 0.004_{\text{us}}=0.17 (5) ,
\ee
or 
\be
N_{3/2}^f=0.35 \pm 0.1_{\text{scale}} \pm 0.04_{\text{NNLO}} \pm 0.05_{1/n} \pm 0.008_{\text{us}}=0.35 (11) ,
\ee
where the latter one is the result of the normalization for the force and can be compared with eq.~(4.4) in Ref.~\cite{Ayala:2020odx},
which reads $0.37(17)$.
The error analysis is also parallel to Ref~\cite{Ayala:2020odx} (we assume symmetric errors in the first place  though)
and the last error shown by ``$\text{us}$"
shows the error associated with the ultrasoft contribution.
In our analysis using Method B', the central value is extracted at $\mu r=1.82$ [or $L=\log(\mu^2 r^2)=1.20$
(see Fig.~\ref{fig4})]
while in the analysis in Ref.~\cite{Ayala:2020odx} the central value is extracted at the minimal sensitivity scale $\mu r=1.52$.

\section{Conclusions and discussion}
\label{sec:5}
In this paper we gave a brief review of the current understanding of 
the renormalons at $u=1/2$ and $u=3/2$ of the static QCD potential in
coordinate and momentum spaces.
We also reconsidered estimation of the normalization constant of the $u=3/2$ renormalon \cite{Sumino:2020mxk,Ayala:2020odx}.
We examined the efficiency of different estimation methods based on a model-like all order series.
Our study agrees with the statement in \cite{Bali:2013pla,Ayala:2014yxa,Ayala:2020odx}
that Method B is superior to Method A.
To improve the estimate further, we proposed to use the consistent scale with 
an asymptotic behavior of perturbative coefficients, instead of the minimal sensitivity scale.
We call it Method B'.
As far as we tested, the proposed method gives most stable result and is most efficient, 
in particular in the sense that it basically makes the error smaller monotonically 
as the order of perturbation theory is raised.

We did not mention the complexity caused by IR divergences 
in perturbative coefficients \cite{Appelquist:1977es,Brambilla:1999qa,Kniehl:1999ud,Brambilla:1999xf} in this paper.
However, related to this, it was pointed out in Ref.~\cite{Sumino:2020mxk} that an unfamiliar renormalon
may arise at $u=1/2$, whose uncertainty is specified as $\sim \LMS r^2 \Delta V^2$.
Also ways to renormalize these IR divergences consistently with the renormalon uncertainties are
discussed therein. These issues need to be further investigated for more precise understanding of renormalons
in the static QCD potential.

Finally we briefly  mention renormalon subtraction methods.
Although we did not mention how one can cope with renormalon uncertainties in this paper,
methods to subtract renormalon uncertainties are being developed
\cite{Lee:2002sn,Lee:2003hh,Ayala:2019uaw,Takaura:2020byt,Ayala:2020odx}.
Recently, a new method has been proposed \cite{Hayashi:2020ylq},
which uses the mechanism of renormalon suppression in momentum space.
We argued in Sec.~\ref{sec:3} that renormalons vanish or are fairly suppressed in momentum space. 
Using this mechanism one can largely suppress renormalons of a general physical observable by
considering Fourier transform to fictional ``momentum space" \cite{Hayashi:2020ylq}.
Higher order computation combined with renormalon subtraction
will be an important direction to give more accurate QCD predictions.

\begin{acknowledgement}
The author is grateful to Yukinari Sumino as 
this work is largely based on Ref.~\cite{Sumino:2020mxk},
which is done in collaboration with him.
This work was supported by JSPS Grant-in-Aid for Scientific Research Grant Number JP19K14711.
\end{acknowledgement}

\appendix
\section{Notation and basic relations}
\label{AppA}
In this appendix we summarize basic knowledge on renormalon
and clarify the notation used in this paper.
The beta function is given by
\be
\mu^2 \frac{d \alpha_s(\mu)}{d \mu^2}=\beta(\alpha_s)
=-b_0 \alpha_s^2-b_1 \alpha_s^3-\cdots .
\ee
The QCD dynamical scale in the $\overline{\rm MS}$ scheme is defined by
\be
\LMS^2/\mu^2
=\exp\lt[ -\lt(\frac{1}{b_0 \alpha_s^2(\mu)}+\frac{b_1}{b_0^2} \log(b_0 \alpha_s(\mu))
+\int_0^{\alpha_s(\mu)} dx \, \lt(\frac{1}{\beta(x)}+\frac{1}{b_0 x^2}-\frac{b_1}{b_0^2 x} \rt) \rt) \rt] . \label{LMS}
\ee

We denote the dimensionless static QCD potential by $v(r)$,  
\be
v(r)=r V_{S}(r)=\sum_{n=0}^{\infty} d^v_n (\mu r) \alpha_s^{n+1}(\mu) ,
\ee
and the dimensionless QCD force by $f(r)$,
\be
f(r)=r^2 \frac{d V_{S}}{d r}=2 \frac{d v}{d L}-v=\sum_{n=0}^{\infty} d_n^f (\mu r) \alpha_s^{n+1}(\mu) ,
\ee 
where $L=\log(\mu^2 r^2)$.
We define the Borel transform of such a perturbative series by
\be
B_X(t) :=\sum_{n=0}^{\infty} \frac{d^X_n(\mu r)}{n!} t^n .
\ee
where $X$ is $v(r)$ or $f(r)$ (or momentum-space potential $\alpha_V(q)$).
Around the singularity at $t=u_*/b_0>0$, it behaves as
\be
B_X(t)=(\mu^2 r^2)^{u_*} \frac{N_{u_*}}{(1-b_0 t /u_*)^{1+\nu_{u_*}}} 
\sum_{k=0}^{\infty} c_{k, u_*}(\mu r) \lt(1-\frac{b_0 t}{u_*} \rt)^k 
+\cdots,  \quad{}  (c_0=1) \label{BorelAroundSing}
\ee
where $N_{u_*}$, $\nu_{u_*}$, and $c_{k, u_*}$ are parameters,
and $\cdots$ denotes a regular function at $t=u_*/b_0$.
The asymptotic behavior of the perturbative coefficient due to the first IR renormalon $t=u_*/b_0$
follows from the above singular Borel transform as
\begin{align}
&d_n^{u_* ({\rm asym})}=N_{u_*} (\mu^2 r^2)^{u_*} \frac{\Gamma(n+1+\nu_{u_*})}{\Gamma(1+\nu_{u_*})}  \lt(\frac{b_0}{u_*} \rt)^n 
\sum_{k=0}^{\infty} c_{k, u_*}(\mu r) \frac{\nu_{u_*} (\nu_{u_*}-1) \cdots (\nu_{u_*}-k+1)}{(n+\nu_{u_*})(n+\nu_{u_*}-1)\cdots (n+\nu_{u_*}-k+1)}.  \label{asymptotic} 
\end{align}
The renormalon uncertainty of $X$ is defined by 
the imaginary part of a regularized Borel integral:
\begin{align}
&{\rm Im} X_{\pm}
={\rm Im} \int_{0\pm i0}^{\infty \pm i0} dt \, B_X(t) e^{-t/\alpha_s(\mu)} \non
&=\pm \frac{\pi}{b_0} \frac{(\mu^2 r^2)^{u_*} N_{u_*}}{\Gamma(1+\nu_{u_*})} u_*^{1+\nu_{u_*}} e^{-\frac{u_*}{b_0 \alpha_s(\mu)}} (b_0 \alpha_s(\mu))^{-\nu_{u_*}} 
\sum_k \nu_{u_*} (\nu_{u_*}-1) \cdot \cdots \cdot (\nu_{u_*}-k+1) \lt(b_0/u_* \rt)^{k} c_{k, u_*}(\mu r) \alpha_s^k(\mu) .
\end{align}
This is renormalization scale independent.
Writing the renormalon uncertainty as
\be
{\rm Im} \, X_{\pm}=\pm K_{u_*} e^{-u_*/(b_0 \alpha_s(r^{-1}))} (b_0 \alpha_s(r^{-1}))^{-\nu_{u_*}} \sum_{k=0}^{\infty} s_{k, u_*} \alpha_s^k(r^{-1}) \label{renormalonuncertainty}
\ee
with $s_0=1$, we have the following relations,
\be
K_{u_*}=\frac{\pi}{b_0} \frac{N_{u_*} }{\Gamma(1+\nu)} u_*^{1+\nu_{u_*}}, \label{relation}
\ee
and
\be
s_{k, u_*}=\nu_{u_*} (\nu_{u_*}-1) \cdot \cdots \cdot (\nu_{u_*}-k+1) (b_0/u_*)^k c_k(\mu r=1) \quad {\rm for}\quad{} k\geq 1 .
\ee

As discussed in Sec.~\ref{sec:2},
since the renormalon uncertainties in coordinate space are given by
\be
K_{u_*} (\LMS^2 r^2)^{u_*} [1+\mathcal{O}(\alpha_s^2(r^{-1}))] ,
\ee
(where $\mathcal{O}(\alpha_s^2)$ can be zero)
one can see that $\nu_{u_*}$ in eq.~\eqref{renormalonuncertainty}
is given by
\be
\nu_{u_*}=u_* b_1/b_0^2
\ee
for $u_*=1/2$ or $3/2$ (where eq.~\eqref{LMS} is used).
One can also calculate $s_{k, u_*}$ and thus $c_{k, u_*}$ by expanding eq.~\eqref{onehalf} or \eqref{threehalf} in $\alpha_s$.

\bibliographystyle{utphys}
\bibliography{BibQCD}

\providecommand{\href}[2]{#2}\begingroup\raggedright\begin{thebibliography}{10}

\bibitem{Appelquist:1977tw}
T.~Appelquist, M.~Dine, and I.~J. Muzinich, ``{The Static Potential in Quantum
  Chromodynamics},''
\href{http://dx.doi.org/10.1016/0370-2693(77)90651-7}{{\em Phys. Lett.}
  {\bfseries 69B} (1977) 231--236}.

\bibitem{Fischler:1977yf}
W.~Fischler, ``{Quark - anti-Quark Potential in QCD},''
  \href{http://dx.doi.org/10.1016/0550-3213(77)90026-8}{{\em Nucl. Phys. B}
  {\bfseries 129} (1977) 157--174}.

\bibitem{Peter:1996ig}
M.~Peter, ``{The Static quark - anti-quark potential in QCD to three loops},''
  \href{http://dx.doi.org/10.1103/PhysRevLett.78.602}{{\em Phys. Rev. Lett.}
  {\bfseries 78} (1997) 602--605},
  \href{http://arxiv.org/abs/hep-ph/9610209}{{\ttfamily arXiv:hep-ph/9610209}}.

\bibitem{Peter:1997me}
M.~Peter, ``{The Static potential in QCD: A Full two loop calculation},''
  \href{http://dx.doi.org/10.1016/S0550-3213(97)00373-8}{{\em Nucl. Phys. B}
  {\bfseries 501} (1997) 471--494},
  \href{http://arxiv.org/abs/hep-ph/9702245}{{\ttfamily arXiv:hep-ph/9702245}}.

\bibitem{Schroder:1998vy}
Y.~Schroder, ``{The Static potential in QCD to two loops},''
  \href{http://dx.doi.org/10.1016/S0370-2693(99)00010-6}{{\em Phys. Lett. B}
  {\bfseries 447} (1999) 321--326},
  \href{http://arxiv.org/abs/hep-ph/9812205}{{\ttfamily arXiv:hep-ph/9812205}}.

\bibitem{Smirnov:2008pn}
A.~V. Smirnov, V.~A. Smirnov, and M.~Steinhauser, ``{Fermionic contributions to
  the three-loop static potential},''
  \href{http://dx.doi.org/10.1016/j.physletb.2008.08.070}{{\em Phys. Lett. B}
  {\bfseries 668} (2008) 293--298},
  \href{http://arxiv.org/abs/0809.1927}{{\ttfamily arXiv:0809.1927 [hep-ph]}}.

\bibitem{Anzai:2009tm}
C.~Anzai, Y.~Kiyo, and Y.~Sumino, ``{Static QCD Potential at Three-Loop
  Order},'' \href{http://dx.doi.org/10.1103/PhysRevLett.104.112003}{{\em Phys.
  Rev. Lett.} {\bfseries 104} (2010) 112003},
\href{http://arxiv.org/abs/0911.4335}{{\ttfamily arXiv:0911.4335 [hep-ph]}}.

\bibitem{Smirnov:2009fh}
A.~V. Smirnov, V.~A. Smirnov, and M.~Steinhauser, ``{Three-Loop Static
  Potential},'' \href{http://dx.doi.org/10.1103/PhysRevLett.104.112002}{{\em
  Phys. Rev. Lett.} {\bfseries 104} (2010) 112002},
\href{http://arxiv.org/abs/0911.4742}{{\ttfamily arXiv:0911.4742 [hep-ph]}}.

\bibitem{Lee:2016cgz}
R.~N. Lee, A.~V. Smirnov, V.~A. Smirnov, and M.~Steinhauser, ``{Analytic
  Three-Loop Static Potential},''
  \href{http://dx.doi.org/10.1103/PhysRevD.94.054029}{{\em Phys. Rev.}
  {\bfseries D94} no.~5, (2016) 054029},
\href{http://arxiv.org/abs/1608.02603}{{\ttfamily arXiv:1608.02603 [hep-ph]}}.

\bibitem{Lee:2002sn}
T.~Lee, ``{Surviving the Renormalon in Heavy Quark Potential},''
  \href{http://dx.doi.org/10.1103/PhysRevD.67.014020}{{\em Phys. Rev.}
  {\bfseries D67} (2003) 014020},
\href{http://arxiv.org/abs/hep-ph/0210032}{{\ttfamily arXiv:hep-ph/0210032
  [hep-ph]}}.

\bibitem{Lee:2003hh}
T.~Lee, ``{Heavy quark mass determination from the quarkonium ground state
  energy: A Pole mass approach},''
  \href{http://dx.doi.org/10.1088/1126-6708/2003/10/044}{{\em JHEP} {\bfseries
  10} (2003) 044},
\href{http://arxiv.org/abs/hep-ph/0304185}{{\ttfamily arXiv:hep-ph/0304185
  [hep-ph]}}.

\bibitem{Ayala:2019uaw}
C.~Ayala, X.~Lobregat, and A.~Pineda, ``{Superasymptotic and hyperasymptotic
  approximation to the operator product expansion},''
  \href{http://dx.doi.org/10.1103/PhysRevD.99.074019}{{\em Phys. Rev.}
  {\bfseries D99} no.~7, (2019) 074019},
\href{http://arxiv.org/abs/1902.07736}{{\ttfamily arXiv:1902.07736 [hep-th]}}.

\bibitem{Takaura:2020byt}
H.~Takaura, ``{Formulation for renormalon-free perturbative predictions beyond
  large-$\beta_0$ approximation},''
  \href{http://dx.doi.org/10.1007/JHEP10(2020)039}{{\em JHEP} {\bfseries 10}
  (2020) 039}, \href{http://arxiv.org/abs/2002.00428}{{\ttfamily
  arXiv:2002.00428 [hep-ph]}}.

\bibitem{Sumino:2020mxk}
Y.~Sumino and H.~Takaura, ``{On renormalons of static QCD potential at $u=1/2$
  and $3/2$},'' \href{http://dx.doi.org/10.1007/JHEP05(2020)116}{{\em JHEP}
  {\bfseries 05} (2020) 116}, \href{http://arxiv.org/abs/2001.00770}{{\ttfamily
  arXiv:2001.00770 [hep-ph]}}.

\bibitem{Ayala:2020odx}
C.~Ayala, X.~Lobregat, and A.~Pineda, ``{Determination of $\alpha(M_z)$ from an
  hyperasymptotic approximation to the energy of a static quark-antiquark
  pair},'' \href{http://dx.doi.org/10.1007/JHEP09(2020)016}{{\em JHEP}
  {\bfseries 09} (2020) 016}, \href{http://arxiv.org/abs/2005.12301}{{\ttfamily
  arXiv:2005.12301 [hep-ph]}}.

\bibitem{Aglietti:1995tg}
U.~Aglietti and Z.~Ligeti, ``{Renormalons and confinement},''
  \href{http://dx.doi.org/10.1016/0370-2693(95)01234-2}{{\em Phys. Lett. B}
  {\bfseries 364} (1995) 75},
  \href{http://arxiv.org/abs/hep-ph/9503209}{{\ttfamily arXiv:hep-ph/9503209}}.

\bibitem{Pineda:1998id}
A.~Pineda, ``{Heavy quarkonium and nonrelativistic effective field theories},''
  ph.d. thesis, 1998.

\bibitem{Hoang:1998nz}
A.~H. Hoang, M.~C. Smith, T.~Stelzer, and S.~Willenbrock, ``{Quarkonia and the
  pole mass},'' \href{http://dx.doi.org/10.1103/PhysRevD.59.114014}{{\em Phys.
  Rev.} {\bfseries D59} (1999) 114014},
\href{http://arxiv.org/abs/hep-ph/9804227}{{\ttfamily arXiv:hep-ph/9804227
  [hep-ph]}}.

\bibitem{Beneke:1998rk}
M.~Beneke, ``{A Quark Mass Definition Adequate for Threshold Problems},''
  \href{http://dx.doi.org/10.1016/S0370-2693(98)00741-2}{{\em Phys. Lett.}
  {\bfseries B434} (1998) 115--125},
\href{http://arxiv.org/abs/hep-ph/9804241}{{\ttfamily arXiv:hep-ph/9804241
  [hep-ph]}}.

\bibitem{Sumino:2014qpa}
Y.~Sumino, ``{Understanding Interquark Force and Quark Masses in Perturbative
  QCD},''
\newblock 2014.
\newblock \href{http://arxiv.org/abs/1411.7853}{{\ttfamily arXiv:1411.7853
  [hep-ph]}}.

\bibitem{Brambilla:1999xf}
N.~Brambilla, A.~Pineda, J.~Soto, and A.~Vairo, ``{Potential NRQCD: an
  Effective Theory for Heavy Quarkonium},''
  \href{http://dx.doi.org/10.1016/S0550-3213(99)00693-8}{{\em Nucl. Phys.}
  {\bfseries B566} (2000) 275},
\href{http://arxiv.org/abs/hep-ph/9907240}{{\ttfamily arXiv:hep-ph/9907240
  [hep-ph]}}.

\bibitem{Lee:1996yk}
T.~Lee, ``{Renormalons beyond one loop},''
  \href{http://dx.doi.org/10.1103/PhysRevD.56.1091}{{\em Phys. Rev.} {\bfseries
  D56} (1997) 1091--1100},
\href{http://arxiv.org/abs/hep-th/9611010}{{\ttfamily arXiv:hep-th/9611010
  [hep-th]}}.

\bibitem{Bali:2013pla}
G.~S. Bali, C.~Bauer, A.~Pineda, and C.~Torrero, ``{Perturbative expansion of
  the energy of static sources at large orders in four-dimensional SU(3) gauge
  theory},'' \href{http://dx.doi.org/10.1103/PhysRevD.87.094517}{{\em Phys.
  Rev. D} {\bfseries 87} (2013) 094517},
  \href{http://arxiv.org/abs/1303.3279}{{\ttfamily arXiv:1303.3279 [hep-lat]}}.

\bibitem{Ayala:2014yxa}
C.~Ayala, G.~Cveti\v~c, and A.~Pineda, ``{The bottom quark mass from the $
  \boldsymbol{\Upsilon} (1S) $ system at NNNLO},''
  \href{http://dx.doi.org/10.1007/JHEP09(2014)045}{{\em JHEP} {\bfseries 09}
  (2014) 045}, \href{http://arxiv.org/abs/1407.2128}{{\ttfamily arXiv:1407.2128
  [hep-ph]}}.

\bibitem{Sumino:2005cq}
Y.~Sumino, ``{Static QCD Potential at $r<\Lambda_\mathrm{QCD}^{-1}$:
  Perturbative Expansion and Operator-Product Expansion},''
  \href{http://dx.doi.org/10.1103/PhysRevD.76.114009}{{\em Phys. Rev.}
  {\bfseries D76} (2007) 114009},
\href{http://arxiv.org/abs/hep-ph/0505034}{{\ttfamily arXiv:hep-ph/0505034
  [hep-ph]}}.

\bibitem{Appelquist:1977es}
T.~Appelquist, M.~Dine, and I.~Muzinich, ``{The Static Limit of Quantum
  Chromodynamics},'' \href{http://dx.doi.org/10.1103/PhysRevD.17.2074}{{\em
  Phys. Rev. D} {\bfseries 17} (1978) 2074}.

\bibitem{Brambilla:1999qa}
N.~Brambilla, A.~Pineda, J.~Soto, and A.~Vairo, ``{The Infrared behavior of the
  static potential in perturbative QCD},''
  \href{http://dx.doi.org/10.1103/PhysRevD.60.091502}{{\em Phys. Rev.}
  {\bfseries D60} (1999) 091502},
\href{http://arxiv.org/abs/hep-ph/9903355}{{\ttfamily arXiv:hep-ph/9903355
  [hep-ph]}}.

\bibitem{Kniehl:1999ud}
B.~A. Kniehl and A.~A. Penin, ``{Ultrasoft effects in heavy quarkonium
  physics},'' \href{http://dx.doi.org/10.1016/S0550-3213(99)00564-7}{{\em Nucl.
  Phys. B} {\bfseries 563} (1999) 200--210},
  \href{http://arxiv.org/abs/hep-ph/9907489}{{\ttfamily arXiv:hep-ph/9907489}}.

\bibitem{Hayashi:2020ylq}
Y.~Hayashi, Y.~Sumino, and H.~Takaura, ``{New method for renormalon subtraction
  using Fourier transform},'' \href{http://arxiv.org/abs/2012.15670}{{\ttfamily
  arXiv:2012.15670 [hep-ph]}}.

\end{thebibliography}\endgroup

\end{document}